# Code C# for chaos analysis of relativistic many-body systems with reactions

I.V. Grossu[a,*], C. Besliu[a], Al. Jipa[a], E. Stan[b], T. Esanu[a], D. Felea[b], C.C. Bordeianu[a]

[a] *University of Bucharest, Faculty of Physics, Bucharest-Magurele, P.O. Box MG 11, 077125, Romania*

[b] *Institute of Space Sciences, Laboratory of Space Research, Bucharest-Magurele, P.O. Box MG 23, 077125, Romania*

ABSTRACT

In this work we present a reactions module for "Chaos Many-Body Engine" (Grossu et al., 2010 [1]). Following our goal of creating a customizable, object oriented code library, the list of all possible reactions, including the corresponding properties (particle types, probability, cross-section, particles lifetime etc.), could be supplied as parameter, using a specific XML input file. Inspired by the Poincare section, we propose also the "Clusterization map", as a new intuitive analysis method of many-body systems. For exemplification, we implemented a numerical toy-model for nuclear relativistic collisions at 4.5 A GeV/c (the SKM200 collaboration). An encouraging agreement with experimental data was obtained for momentum, energy, rapidity, and angular $\pi^-$ distributions.

## Program summary

*Manuscript title: Code C# for chaos analysis of relativistic many-body systems with reactions*
*Catalogue identifier : AEGH_v2_0*
*Authors: I.V.Grossu, C. Besliu, Al. Jipa, E. Stan ,T. Esanu, D. Felea, C.C. Bordeianu*
*Program title: Chaos Many-Body Engine v02*
*Licensing provisions:*
*Programming language: Visual C# .NET 2005*
*Computer(s) for which the program has been designed: PC*
*Operating system(s) for which the program has been designed: .Net Framework 2.0 running on MS Windows*
*RAM required to execute with typical data: 128 Megabytes*
*Has the code been vectorised or parallelized?: each many-body system is simulated on a separate execution thread*
*Number of processors used: one processor for each many-body system*
*Supplementary material: Windows forms application for testing the engine.*
*CPC Library Classification:6.2, 6.5*
*Catalogue identifier of previous version: AEGH_v1_0*
*Journal reference of previous version: Computer Physics Communications 181 (2010) 1464–1470*
*Does the new version supersede the previous version?: Yes*
*External routines/libraries used: .Net Framework 2.0 Library*
*Nature of problem: Chaos analysis of three-dimensional, relativistic many-body systems with reactions.*
*Solution method: Second order Runge-Kutta algorithm for simulating relativistic many-body systems with reactions. Object oriented solution, easy to reuse, extend and customize, in any development environment which accepts .Net assemblies or COM components. Treatment of two particles reactions and decays. For each particle, calculation of the time measured in the particle reference frame, according to the instantaneous velocity. Possibility to dynamically add particle properties (spin, isospin etc.), and reactions/decays, using a specific XML input file. Basic support for Monte Carlo simulations. Implementation of: Lyapunov exponent, "fragmentation level", "average system radius", "virial coefficient", "clusterization map", and energy conservation precision test. As an example of use, we implemented a toy-model for nuclear relativistic collisions at 4.5 A GeV/c.*

*Restrictions:*
*Additional comments: Easy copy/paste based deployment method.*
*Running time: quadratic complexity*

**1. Introduction**

In a previous work [1], we presented “Chaos Many-Body Engine” code library for chaos analysis of relativistic many-body systems. Inspired by existing studies on Fermi nuclear systems [2-5], we tried to apply chaos theory to the more complex case of nuclear relativistic collisions at 4.5 A GeV/c. In this domain of energies it is possible to apply the many-body representation, as the nucleon de Broglie wavelength is less than the average inter-nucleonic distance, and the nucleon mean free path is shorter than the target radius [6-10]. On the other hand, the two-nucleon reactions are a key point in understanding the processes involved in nuclear collisions. In this context we motivate our efforts of developing a C# [11] reactions module, integrated with “Chaos Many-Body Engine”.

**2. Two particles reactions and decays**

The second version of “Chaos Many-Body Engine” library is designed as a general numerical solution for the simulation of three-dimensional relativistic many-body systems with reactions. We implemented: two particles reactions, decays, and stimulated decays.

**a+b->c+d reactions.** For simplicity, we considered that only the momentum component oriented towards the direction between the two bodies is changing during the collision. Working in a Cartesian system in which this direction corresponds to abscise, by applying the relativistic energy and momentum conservation laws, wrote in the natural system of unities, we obtained the following second degree equation:

$$(\propto^2 - 1)E_c^2 + 2 \propto \beta E_c + \beta^2 + p_{ya}^2 + p_{za}^2 + m_c^2 = 0 \quad (1)$$

with:

$$\alpha = \frac{E_a + E_b}{p_{xa} + p_{xb}} \quad (2)$$

$$\beta = \frac{m_d^2 + p_{yb}^2 + p_{zb}^2 - (E_a + E_b)^2 + (p_{xa} + p_{xb})^2 - p_{ya}^2 - p_{zb}^2 - m_c^2}{2(p_{xa} + p_{xb})} \quad (3)$$

where, $m_a$, $m_b$, $m_c$, $m_d$ are the rest masses, $p_{xa}$, $p_{xb}$, $p_{xc}$, $p_{xd}$, $p_{ya}$, $p_{yb}$, $p_{yc}$, $p_{yd}$, $p_{za}$, $p_{zb}$, $p_{zc}$, $p_{zd}$ the Cartesian components of momentum, and $E_a$, $E_b$, $E_c$, $E_d$ are the particle energies.

The momentums of the two resultant particles are given by:

$$p_c = \pm\sqrt{E_c^2 - m_c^2} \quad (4)$$

$$\vec{p}_d = \vec{p}_a + \vec{p}_b - \vec{p}_c \quad (5)$$

Only the physical solutions of equations (1-5) are chosen. We impose also two additional conditions:

- The two colliding particles must approach each other

- The distance between the two particles must be less than a specific value, $r_{max}$, calculated in agreement with the cross section $\sigma$:

$$r_{max} = \sqrt{\frac{\sigma}{\pi}} \quad (6)$$

**a->c+d decays.** For simplicity, we assumed that each decay take place after the corresponding lifetime, measured in particle's own reference system. Thus, the program was improved in order to adjust each particle lifetime, according to the corresponding instantaneous velocity:

$$t_2 = t_1 + \frac{dt}{\sqrt{1 - \beta(t_1)^2}} \quad (7)$$

where $t$ is the time, $\beta = v/c$, and $dt$ is the integration interval.

In the frame of decaying particle, the two new particles are emitted on a random direction, with the following momentum:

$$p = p_c = p_d = \frac{1}{2m_a}\sqrt{[m_a^2 - (m_c + m_d)^2][m_a^2 - (m_c - m_d)^2]} \quad (8)$$

For obtaining the result in the simulation reference frame, we must relativistic compose the two calculated velocities, with the transport velocity.

**Stimulated decays** are decays triggered by collisions. At least one participant must have at least one decay schema. This kind of reactions is implemented by simply setting the particles lifetime values to zero on collision events.

It is important to notice that more complex schemas could be considered by employing various combinations of the, previous described, basic reactions. For example, for a + b -> c + d + e, one can use a virtual particle "f", with zero lifetime, which decays into "d" and "e". Thus, the engine will process the following sequence: a + b -> c + f ; f -> d + e.

## 3. Program description

Our main goal was to create a highly configurable reactions module, integrated with "Chaos Many Body Engine". Thus, the list of all reaction schemas which applies to the many-body system of interest is stored in a *DsReactions* typed dataset object. The main advantages of using datasets [12] are related to XML serialization, and very easy filtering capabilities, based on SQL-like query strings.

The *DsReactions* dataset contains three datatables. The *DsReactions.Particle* table is used for defining all necessary particle types. It contains the unique identifier (*IdParticle*), and the most common particle properties (mass, charge, and lifetime). The *DsReactions.ParticleProperties* datatable is used for dynamically defining particle properties (spin, isospin, color charge and so on). It is in a many to one relation with *DsReactions.Particle*, and contains each property name and value. One new *Particle* class instance can be created, based on information stored in the *DsReactions.Particle* and *DsReactions.ParticleProperties* datatables, using the *NBody.NewParticle* method, which expects the particle type (*IdParticle*) as parameter.

The *DsReactions.Reactions* datatable is used for defining the list of all possible reactions for the specific system of interest. The *IdParticleIn01* and *IdParticleIn02* columns represent the unique identifiers of the input particles, while *IdParticleOut01* and *IdParticleOut02* are the reaction output identifiers. The *MaxDistance* column is directly related to the cross-section parameter (Eq. (6)). For avoiding also any unwanted effect related to infinity values near origin, we considered a minimum distance (the *MinDistance* column), under which the reaction is forbidden. For decays, this property has a different signification. It represents the initial distance between the two, new generated, particles. When more channels are possible, the reaction probability can also be specified. The same datatable is used for all reaction types. Thus, a decay is specified by setting *IdParticleIn02=null*, while for a stimulated decay, one will set both *IdParticleOut01*, and *IdParticleOut02* to *null*.

The list of all allowed reactions can be supplied/modified by simply creating/changing the XML file used as data source for the *DsReactions* dataset. The user should carefully define each reaction, in agreement with all specific conservation laws (charge, baryonic number and so on), and avoiding also circular references.

Based on a second order Runge-Kutta algorithm, the *Nbody.Next* method is used for "moving" the many-body system into its next status [1]. We improved this method for finding also all particles susceptible to participate to reactions. Thus, we fill one generic list of *Reaction* objects with all constituent pairs for which the mutual distance is less than the maximum cross-section radius, Eq. (6), and another one, with all particles for which the own time, measured in the own reference system, is greater or equal with the corresponding lifetime. The two lists are used as input for the reactions engine. For optimization reasons, the reactions module is called with a frequency which can be specified as parameter. It is important to notice that the program will automatically increase this frequency in certain circumstances (the number of particles susceptible to participate to reactions changed, and/or, at least one reaction took place in the current iteration).

The *ReactionEngine* class contains the main processing routines. The *GetReactions* method assures that one constituent will participate to only one reaction. The decays are treated in priority. The *CheckReaction* method is used for finding all reactions which apply to the input data (the, previous described, generic lists), according to the information stored in the *DsReactions* dataset. The *GetReactionOutput* and *GetDecayOutput* methods select only the physical solutions which satisfy the energy and momentum conservation laws (equations (1-5)). If there are more possible reaction schemas for the same input, one solution is randomly chosen, according to its probability (the *DsReactions.Reactions.Probability* column).

The *Nbody.ProcessReactions* function is used for applying the processing result to the *NBody* instance. It removes the old particles, and adds the new ones from/to the *mParticles* generic list. The *OnReactions* event is raised if at least one reaction took place. As the collisions are inelastic, the kinetic and potential energies are also adjusted after each reaction event, in order to avoid any impact on the energy conservation precision test [13]:

$$-log_{10}\left|\frac{E(t)-E(t=0)}{E(t=0)}\right| \quad (9)$$

The structure of the simulation output files was changed according to the new necessities, trying to assure also the compatibility with the previous version. Thus, in the header file we added the creation time of each particle (the *Time* column), an integer (*Particle*) which uniquely identifies each particle instance, and the particle type (*Id*), which corresponds to the *IdParticle* column defined in the *DsReactions* dataset. As the composition of the system could change in time, the unique instance identifier (*Particle*) was added also to the data file.

**5. Chaos analysis of many-body systems**

In [1] we implemented the multi-dimensional Lyapunov Exponent [14,15]:

$$L \stackrel{\text{def}}{=} \lim_{t\to\infty} \frac{1}{t} ln \frac{d(t)}{d(0)} = \lim_{t\to\infty} L(t) \quad (10)$$

where *d(t)* represents the phases space distance between the two systems:

$$d = \sqrt{\sum_{i=1}^{n} ((\vec{r}_{i1} - \vec{r}_{i2})^2 + (\vec{p}_{i1} - \vec{p}_{i2})^2)} \quad (11)$$

where $r_{i1}$ , $p_{i1}$, are the position, respectively the momentum, of the particle "*i*" from the first system, and $r_{i2}$ , $p_{i2}$, are the analogous coordinates of the second system.

As the structure of one many-body system with reactions could change in time, the previous distance cannot be applied any more. We propose instead a "global" approach:

$$d = \sqrt{\left(\sum_{i=1}^{n} \vec{r}_{i1} - \sum_{i=1}^{n} \vec{r}_{i2}\right)^2 + \left(\sum_{i=1}^{n} \vec{p}_{i1} - \sum_{i=1}^{n} \vec{p}_{i2}\right)^2} \quad (12)$$

The *BaseSystemData.GlobalLyapunovExponent* static method is implementing the Lyapunov Exponent based on this distance definition.

Inspired by the Poincare section [15,16], we propose also a new intuitive investigation method. Thus, for one many-body constituent we considered the following set:

$$M_i = \left\{(x_i(t), y_i(t)) | \frac{dp_{xi}}{dt}(t) = 0 \vee \frac{dp_{yi}}{dt}(t) = 0 \vee \frac{dp_{zi}}{dt}(t) = 0\right\} \quad (13)$$

We defined the "Clusterization Map" as:

$$M = \bigcup_{i=1}^{n} M_i \quad (14)$$

Using this technique, one obtains a "bi-dimensional projection" of all clusters created inside the system. For a better precision, the *BaseSystemData.ClusterizationMap* static method is searching all points for which the derivatives of momentum components signs are changing.

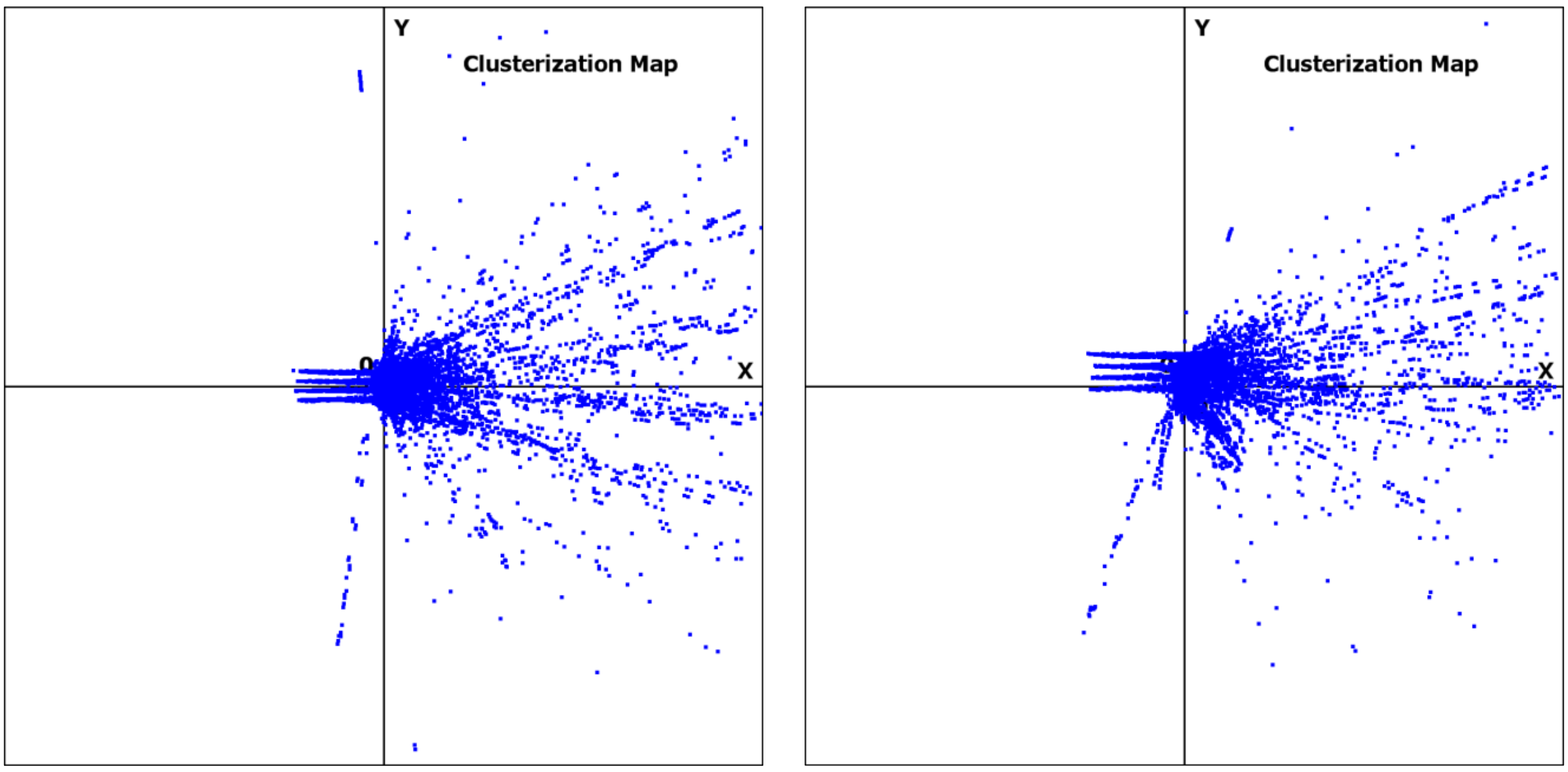

**Fig.1.** Two Clusterization Map examples for Cu+Cu at 4.5 A GeV/c nuclear collision simulation, obtained with Chaos Many-Body. (Left) Central collision. (Right) Non-central collision.

### 6. Application to the C-C nuclear collision at 4.5 A GeV/c

As an example of use, we implemented a simplified model for nuclear collisions at 4.5 A GeV/c (the SKM200 collaboration [8,17-19]). The colliding nuclei are represented as two distinct sets of nucleons, initially placed in the vertices of a cubical network, with the edge calculated in agreement with the nucleus radius $r=r_0A^{1/3}$. The target is initially at rest, while the incident momentum of the projectile constituents could be specified as parameter. For simplicity, the Lorenz contraction is ignored.

We employed a finite depth Yukawa potential well, together with a coulombian, and a short distance repulsive term:

$$V(r_{ij}) = \begin{cases} -k\,ln(r), & r_{ij} \le 1.134Fm \\ -V_0 \dfrac{e^{-\frac{r_{ij}}{a}}}{\frac{r_{ij}}{a}} + \dfrac{q_i q_j}{4\pi\epsilon r_{ij}} - k\,ln(r), & 1.134Fm < r_{ij} < 2.2Fm \\ -V_0 \dfrac{e^{-\frac{r_{ij}}{a}}}{\frac{r_{ij}}{a}} + \dfrac{q_i q_j}{4\pi\epsilon r_{ij}}, & r_{ij} \ge 2.2Fm \end{cases} \quad (9)$$

where $V_0=35MeV$ is the depth and $a=2Fm$ is the radius of the potential well, $r_{ij}$ represents the distance between the two bodies, and $q$ is the electric charge. We empirically chose $k=200$.

The following interactions were considered:

$$p + n \rightarrow p + \Delta^0 \qquad 50\%$$
$$p + n \rightarrow \Delta^+ + n \qquad 50\%$$
$$p + p \rightarrow \Delta^+ + p \qquad 100\%$$
$$n + n \rightarrow \Delta^0 + n \qquad 100\%$$
$$\Delta^0 \rightarrow p + \pi^- \qquad 50\%$$
$$\Delta^0 \rightarrow n + \pi^0 \qquad 50\%$$
$$\Delta^+ \rightarrow p + \pi^0 \qquad 50\%$$
$$\Delta^+ \rightarrow n + \pi^+ \qquad 50\%$$

Working with a $10^{-3}$ Fm/c value for temporal resolution, we simulated 500 C+C events, with random collision parameters. The comparison with experimental data illustrates an encouraging result for momentum, energy, rapidity, and angular $\pi^-$ distributions.

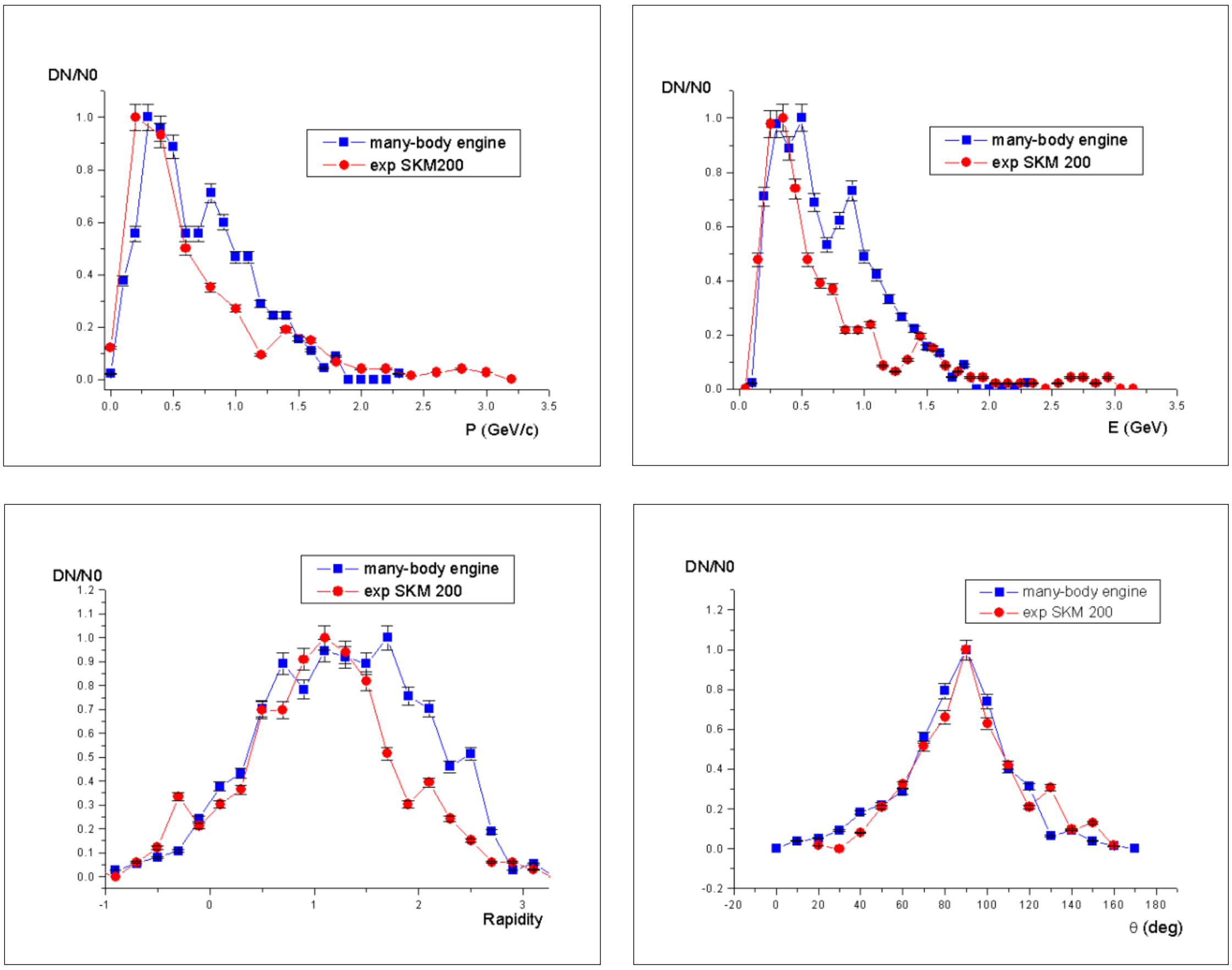

**Fig.2.** Comparison between Chaos Many Body Engine and experimental $\pi^-$ distributions for C+C at 4.5 A GeV/c. (Up Left) momentum distribution. (Up Right) energy distribution. (Down Left) rapidity distribution. (Down Right) angular distribution.

## 7. Conclusion

Starting from exiting analysis of Fermi nuclear systems, one of our main goals was to apply chaos theory to relativistic nuclear collisions at 4.5 A GeV/c (the SKM200 collaboration). In this context, we developed a C# reactions module for "Chaos Many-Body

Engine". An important attention was paid to the application flexibility and extensibility. Thus, we improved the code in order to accept any number of, dynamically loaded, particle properties (spin, isospin, color charge and so on). The list of all particles and possible reactions could be also supplied as a simulation parameter using a specific XML input file.

As the structure of one many-body system with reactions could change in time, in the implementation of the multi-dimensional Lyapunov Exponent we considered a particular definition for the distance between two systems, equation (12). We propose also the "Clusterization Map", equation (14), as a new analysis method, which can bring intuitive information on the clusters created over the time in the system.

The, previous described, example of use (C+C collision) does not require significant resources to run (CPU 1.0GHz, 128M free RAM, .Net Framework 2.0 running on MS Windows XP or later). However, the resources could become a critical problem with the increasing of reactions and particles numbers (e.g. Cu+Cu or Au+Au collisions).

Using a simplified model for C+C nuclear collision at 4.5 A GeV/c, we obtained some encouraging results for momentum, energy, rapidity, and angular $\pi^-$ distributions (Fig.2). Further analysis along those lines is currently in progress.

**References**

[1] I.V. Grossu, C. Besliu, Al. Jipa, C.C. Bordeianu, D. Felea, E. Stan, T. Esanu, Computer Physics Communications 181 (2010) 1464–1470

[2] G.F. Burgio, F.M. Baldo, A. Rapisarda, P. Schuck, Phys. Rev. C 58 (1998) 2821–2830.

[3] D. Felea, PhD thesis, University of Bucharest, Physics Department, 2002.

[4] C.C. Bordeianu, D. Felea, C. Besliu, Al. Jipa, I.V. Grossu, Computer Physics Communications 179 (2008) 199–201.

[5] C.C. Bordeianu, D. Felea, C. Bes_liu, Al. Jipa, I.V. Grossu, Commun Nonlinear Sci Numer Simulat 16 (2011) 324–340

[6] S.Nagamya - Prog.Part.Nucl.Phys.XV(1985)363

[7] R.Stock - Prog.Part.Nucl.Phys.XV(1985)455

[8] Al. Jipa, C. Besliu, Elemente de fizica nucleara relativista – Note de curs, Editura Universitatii din Bucuresti, Bucuresti, Romania, 2002

[9] C. Besliu, Al.Jipa – Rev.Roum.Phys.33(1988) 409

[10] Al.Jipa, PhD thesis, University of Bucharest, Physics Department, 1989

[11] Christian Nagel, Bill Evjen, Jay Glynn, Morgan Skinner, Karli Watson, Professional C# 2008, Wiley, Indianapolis, Indiana, 2008

[12] Glenn Johnson, Programming Microsoft ADO.NET 2.0 Applications: Advanced Topics, Microsoft Press, Washington, US, 2006

[13] R.H. Landau, M.J. Paez, C.C. Bordeianu, Computational Physics: Problem Solving with Computers, Wiley-VCH-Verlag, Weinheim, 2007

[14] M. Sandri, Numerical Calculation of Lyapunov Exponents, Mathematica J.6, 78-84, 1996.

[15] C.C. Bordeianu, D. Felea, C. Besliu, Al. Jipa, I.V. Grossu, Computer Physics Communications 178 (2008) 788-793.

[16] G.P. Williams, Chaos Theory Tamed, Joseph Henry Press, Washington D.C., 1997.

[17] A.U. Abdurakhimov et al, Preprint JINR 13-10692 (1977)

[18] P.Rice-Evans – Spark, Streamer, Proportional and Drift Chambers, The Richelieu Press, London, 1974

[19] C. Besliu, N.Ghiordanescu, M.Pentia – Studii si cercetari de fizica 29 (1977) 817